\documentclass[reprint,superscriptaddress, amsmath,amssymb, aps,floatfix,]{revtex4-1}

\usepackage{graphicx}
\usepackage{dcolumn}
\usepackage{bm}
\usepackage{textcomp}
\usepackage{float} 
\usepackage{xcolor}
\usepackage{graphicx,psfrag}
\usepackage[latin1]{inputenc}
\usepackage{natbib}
\usepackage{amsmath, amsthm, amssymb}
\usepackage{color}
\usepackage{comment}
\usepackage{hyperref}
\usepackage{xcolor}
\usepackage{esvect}

\usepackage{lipsum} 
\usepackage{graphicx} 
\usepackage{environ} 
\usepackage{bm}

\begin{document}

\title{Quantifying Spin-Orbit Torques in Antiferromagnet/Heavy Metal Heterostructures}
\author{Egecan Cogulu}
 \email{egecancogulu@nyu.edu}
\affiliation{Center for Quantum Phenomena, Department of Physics, New York University, NY 10003, USA}
\author{Hantao Zhang}%
 \affiliation{Department of Electrical and Computer Engineering, University of California, Riverside, CA 92521, USA}
\author{Nahuel N. Statuto}%
\affiliation{Center for Quantum Phenomena, Department of Physics, New York University, NY 10003, USA}
\author{Yang Cheng}%
 \affiliation{Department of Physics, The Ohio State University, Columbus, OH 43210, USA}
\author{Fengyuan Yang}%
 \affiliation{Department of Physics, The Ohio State University, Columbus, OH 43210, USA}
\author{Ran Cheng}%
 \affiliation{Department of Electrical and Computer Engineering, University of California, Riverside, CA 92521, USA}
  \affiliation{Department of Physics and Astronomy, University of California, Riverside, CA 92521, USA}
\author{Andrew D. Kent}%
\affiliation{Center for Quantum Phenomena, Department of Physics, New York University, NY 10003, USA}
\date{\today}

\begin{abstract}
The effect of spin currents on the magnetic order of insulating antiferromagnets (AFMs) is of
fundamental interest and can enable new applications. Toward this goal, characterizing the spin-orbit torques (SOT) associated with AFM/heavy metal (HM) interfaces is important. Here we report the full angular dependence of the harmonic Hall voltages in a predominantly easy-plane AFM, epitaxial c-axis oriented $\alpha$-Fe$_2$O$_3$ films, with an interface to Pt. By modeling the harmonic Hall signals together with the $\alpha$-Fe$_2$O$_3$ magnetic parameters, we determine the amplitudes of field-like and damping-like SOT. Out-of-plane field scans are shown to be essential to determining the damping-like component of the torques. In contrast to ferromagnetic/heavy metal heterostructures, our results demonstrate that the field-like torques are significantly larger than the damping-like torques, which we correlate with the presence of a large imaginary component of the interface spin-mixing conductance. Our work demonstrates a direct way of characterizing SOT in AFM/HM heterostructures.
 
\end{abstract}
\maketitle
Recently antiferromagnetic materials have been gathering increasing attention from the spintronics community due to their advantageous properties such as fast spin dynamics, low susceptibility, and magnetic moment compensation~\cite{Gomonay2014,Jungwirth2016,Baltz2018,Fukami2020}. 
Detecting and manipulating antiferromagnetic order electrically is an important milestone for realizing devices based on AFMs~\cite{Wadley2016,Cheng2016,Gray2019,Cheng2020,Cogulu2021}. It is known that spin-orbit torques (SOT) are one of the most effective ways to manipulate magnetic order in both ferromagnets (FM) and ferrimagnets \cite{Gomonay2010,Cheng2014,Avci2014,Bodnar2018,Chen2018,Zhou2018,Parthasarathy2021,Shao2021,Zhu2021}. However, their effectiveness is less well explored and quantified for AFMs, particularly insulating AFMs.  
Therefore, characterizing the type and the amplitude of the SOT is crucial for understanding and predicting AFM dynamics. 
An important technique to characterize SOT is harmonic Hall measurements~\cite{Baldrati2018,Cheng2019}. Although this technique has been used extensively in FM/heavy metal (HM) bilayers, there are few studies of AFM. 

In this work, we use harmonic Hall measurements to characterize the type and amplitude of spin-orbit torques in $\alpha$-Fe$_2$O$_3$/Pt bilayer heterostructures. First, we show the full angular dependence of $1^\mathrm{st}$ and $2^\mathrm{nd}$ harmonic Hall signal in 3 orthogonal planes.
\begin{figure}[ht]
  \centering
   \includegraphics[width=1\columnwidth]{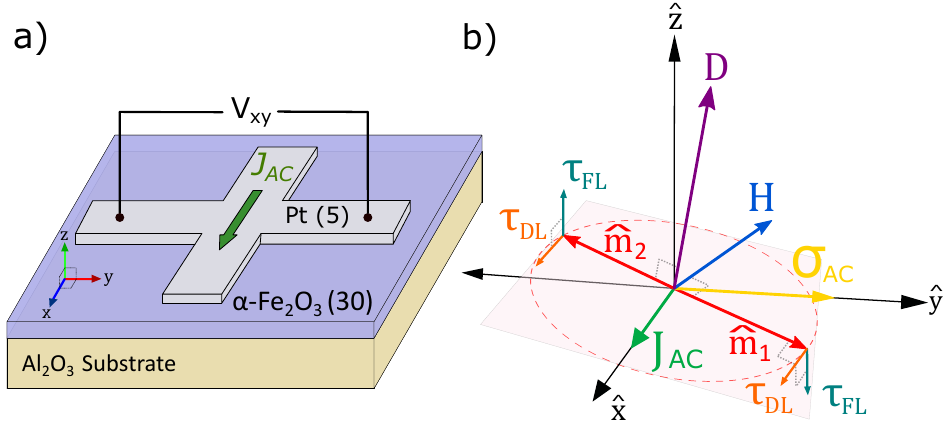}
    \caption{a)	Transverse measurement configuration showing the Pt Hall cross with 5 $\mu$m channel width. An AC current \bm{$J_{\textrm{AC}}$} is applied in the $\hat{x}$-direction, and the transverse voltage $V_{xy}$ is measured with a lock-in amplifier. The thicknesses of the $\alpha$-Fe$_2$O$_3$ layer and Pt layers are 30 nm and 5 nm respectively. b) Modeling geometry. $\bm{D}$ represents both the DMI vector and the hard axis direction that defines the easy plane. $\bm{m}_{1}$ and $\bm{m}_{2}$ are the sublattice moments and $\bm{H}$ is the external magnetic field. The spin accumulation $\bm{\sigma}_\textrm{AC}$ is in the $\hat{y}$-direction. The resulting spin-orbit torques on the sublattice moments are decomposed into field-like ($\tau_{\textrm{FL}}$) and damping-like components ($\tau_{\textrm{DL}}$).
    }
\label{Fig1:Measurement Geometry}
\end{figure}
Then, we develop a model of the response that accounts for the magnetic properties of $\alpha$-Fe$_2$O$_3$, including its magnetic anisotropy, exchange and Dzyaloshinskii-Moriya interactions (DMI), and compare the model with the experimental results. By fitting the data from six measurements together (the $1^\mathrm{st}$ and $2^\mathrm{nd}$ harmonic response with the field rotated in three orthogonal planes), we extract the amplitudes of the damping-like and field-like torques. Surprisingly and contrary to the case of  ferromagnets/heavy metal heterostructures, we find field-like torques to be significantly larger than damping-like torques. We also show that there is an ordinary Hall-effect (OHE) contribution to the response when the applied magnetic field has a component out of the film plane and spin Seebeck effect (SSE) contribution to the in-plane field results. Further, our model suggests a small canting of the easy plane of the AFM. Finally, by performing finite-element simulations, we show that Oersted fields are much smaller than the field-like SOT.

Figure \ref{Fig1:Measurement Geometry}(a) shows a schematic of the experimental configuration for the harmonic Hall measurements. We perform the measurements on 30 nm thick epitaxial c-axis oriented $\alpha$-Fe$_2$O$_3$ grown on single crystalline Al$_2$O$_3$ (0001) substrates at 500\textdegree C using off-axis sputtering~\cite{Cheng2019,Cheng2020}.
The films are capped with 5 nm thick Pt in-situ, deposited at room temperature. Finally, we pattern the Pt layer into a 5 $\mu$m  width and 15 $\mu$m length Hall-cross structure using electron-beam lithography and Ar$^+$ plasma etching. All measurements are done with a Quantum Design DynaCool PPMS system. An AC current with frequency 953 Hz is supplied with a Keithley 6221 current source and detected with a SRS lock-in amplifier SR830; current densities of $4\times10^{9}$~A/m$^2$ and $6\times10^{10}$~A/m$^2$ were used for 1$^\mathrm{st}$ and 2$^\mathrm{nd}$ harmonic measurements, respectively. All experiments were performed at 300 K.

\begin{figure}[t]
  \centering
   \includegraphics[width=1\columnwidth]{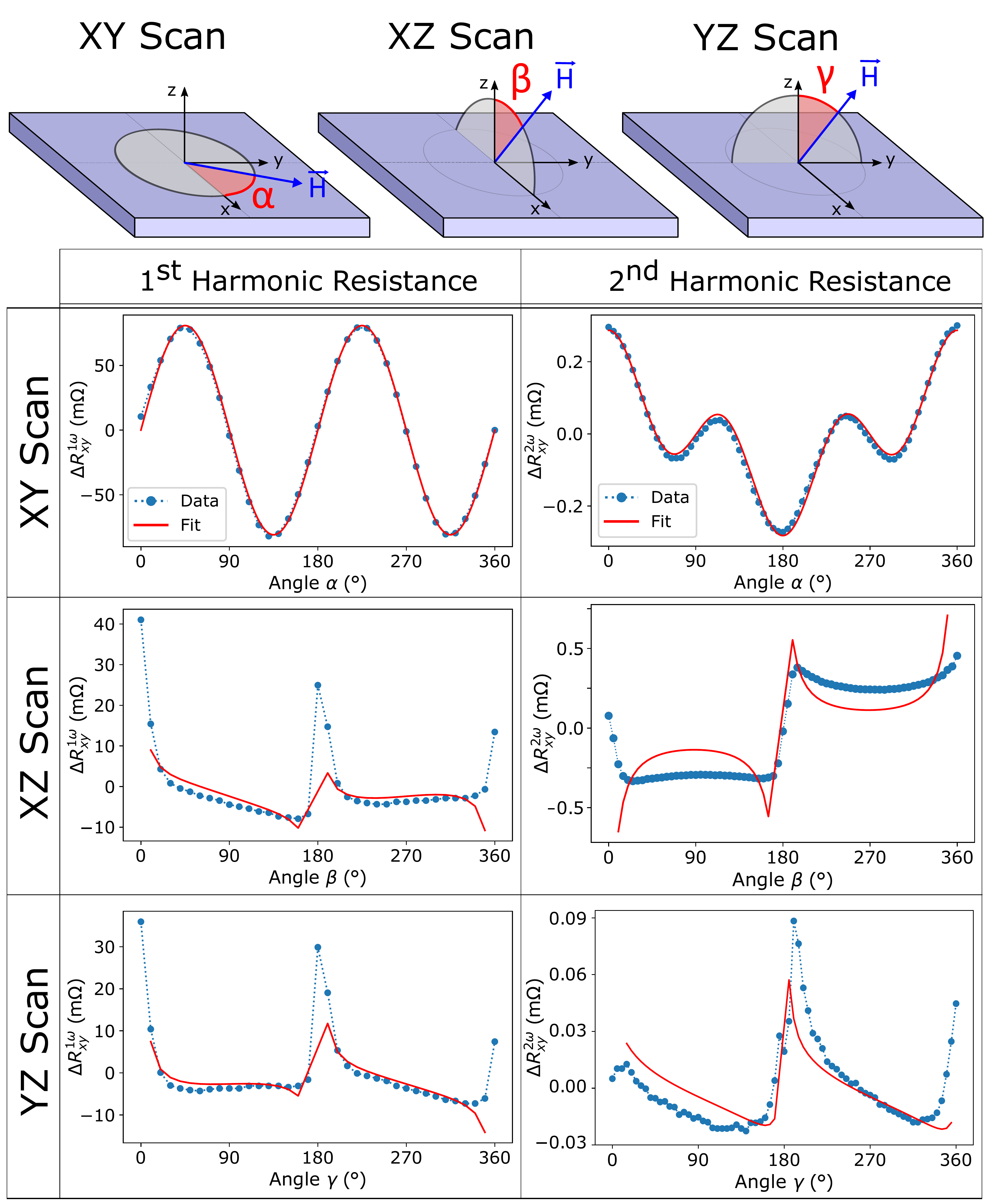}
    \caption{Angular dependence of $1^\mathrm{st}$ and $2^\mathrm{nd}$ harmonic (columns) resistance signals in \textit{XY}, \textit{XZ}  and \textit{YZ} scans (rows) at $\mu_{0}\textrm{H}$ = 3 T. The current amplitudes are 100 $\mu$A and 1.5 mA in the $1^\mathrm{st}$ and $2^\mathrm{nd}$ harmonic measurements, respectively. The blue dots are the data points whereas the red line is the fit resulting from our model. In all six fits the same material parameters are used: $H_{D}$ = 2 T, $H_{K}$ = 0.01 T and H$_{ex}$ = 900 T. From these fits, we extract: $H_{\textrm{FL}}\approx 4.5 \times10^{-3}$ T and $H_{\textrm{DL}}\approx 2.5\times 10^{-5}$ T. In $1^\mathrm{st}$ harmonic \textit{XZ} and \textit{YZ} scans, the fit includes a cosine term to represent an ordinary Hall effect contribution. Similarly, in $2^\mathrm{nd}$ harmonic \textit{XY} scan, a cosine term is added to the fit to represent the SSE contribution. The geometry of magnetic field scans in the 3 principal planes (\textit{XY}, \textit{XZ} and \textit{YZ} scans) and the definition of the angles $\alpha$, $\beta$ and $\gamma$ are shown at the top. The current always flows in the $\hat{x}$ direction and the resulting spin accumulation $\bm{\sigma}_\textrm{AC}$ is in the $\hat{y}$ direction.}
\label{Fig2:Fits}
\end{figure}

Figure \ref{Fig1:Measurement Geometry}(b) shows the geometry of the model we used to explain the experimental data. The antiferromagnet's hard axis and DMI are represented by the same vector $\bm{D}$. The AFM sublattice moment directions are indicated by unit vectors $\bm{m}_1$ and $\bm{m}_2$. The N\'eel vector $\bm{n} = (\bm{m}_{1}-\bm{m}_{2})/2$  lies in the easy plane, which is perpendicular to $\bm{D}$, indicated by the red plane. $\bm{H}$ is the applied magnetic field and $\bm{\sigma}_{\textrm{AC}}$ is the spin accumulation at the interface caused by $\bm{J}_{\textrm{AC}}$. The resulting torques on the sublattice moments are decomposed into field-like ($\tau_{\textrm{FL}}$) and damping-like components ($\tau_{\textrm{DL}}$). Notice we do not constrain $\bm{D}$ to be parallel to the $\hat{z}$-axis, as canting of the easy plane in AFMs, due to the strain effects from the substrate has been reported~\cite{Cogulu2021,Schmitt2021}. As shown at the top of Fig. \ref{Fig2:Fits}, the applied field $\bm{H}$ is rotated in one of the three Cartesian planes: \textit{XY}, \textit{XZ} or \textit{YZ}, where the current $\bm{J}_{\textrm{AC}}$ is in the $\hat{x}$ direction and $\hat{z}$ is perpendicular to the film plane.

Figure \ref{Fig2:Fits} shows the full angular dependence of Hall signal for both harmonics (blue points), for all three field scans, at a selected magnetic field strength of $\mu_{0}H = 3$~T together with the model fit (red lines). \textit{XY}, \textit{XZ}  and \textit{YZ} plane scans of magnetic field and the angles $\alpha$, $\beta$ and $\gamma$ are shown at the top of Fig. \ref{Fig2:Fits}. In the in-plane XY scan, both $1^\mathrm{st}$ and $2^\mathrm{nd}$ harmonic Hall signals follows smooth trigonometric functions as we rotate the field in the sample plane. This is the expected behavior because spin-Hall magnetoresistance (SMR) and field-like torques are the main contributors in $1^\mathrm{st}$ and $2^\mathrm{nd}$ \textit{XY} scan harmonic signals, respectively. On the other hand, for out-of-plane field rotation experiments---\textit{XZ}  and \textit{YZ} field scans---the angular dependencies do not follow a simple trigonometric function. A comprehensive model is needed in order to explain all six scans at the same time, as we discuss below.

In the sample geometry indicated in Fig.~\ref{Fig1:Measurement Geometry}(b), an AC current density $J_\textrm{AC}(t)=J_{0}\cos \omega t$ (with angular frequency $\omega/2\pi= 953$ Hz) applied in the $\hat{x}$ direction generates an oscillating spin accumulation $\sigma_\textrm{AC}(t)$ in the $\hat{y}$ direction, which acts on the $\alpha$-Fe$_2$O$_3$ film through both field-like and damping-like torques. The measured Hall voltage $V_{xy}$ arises from the spin-Hall magnetoresistance (SMR): $V_{xy}(t)= A_{x} J_{0}R_{xy}[\bm{n}(J_\textrm{AC},\bm{H})]\cos \omega t$, where $A_{x}$ is the cross sectional area of Pt perpendicular to the $\hat{x}$ direction and the SMR has the form $R_{xy} \sim n_{x}n_{y}$ \cite{Chen2013,Hayashi2014,Fisher2018,Lebrun2019} (the contribution of $m_{x}m_{y}$ is negligible). Since the current-induced torques only result in a slight deviation of $\bm{n}$ from its equilibrium orientation, we can expand the SMR as:
\begin{align}
R_{xy}[\bm{n}(J_{AC},\bm{H})] &= R_{xy}[\bm{n}(0,\bm{H})] \\& + J_{0}\left( \frac{\partial R_{xy}}{\partial \bm{n}}\right) \frac{\partial \bm{n}}{\partial J_{AC}}\Bigr|_{(0,\bm{H})}\cos \omega t + h.o. \nonumber,
\end{align}
where the first term $R_{xy}[\bm{n}(0,\bm{H})]$ is the unperturbed SMR, the resistance response that is independent of the applied current density, which leads to the first harmonic signal $V_{xy}^{1\omega}(t)=A_{x} J_{0}R_{xy}[\bm{n}(0,\bm{H})]\cos \omega t$. The second term, which is itself proportional to $J_{0}$, gives rise to the second harmonic response  $V_{xy}^{2\omega}(t)$. Therefore, to find the harmonic responses, we need to calculate $\bm{n(0,\bm{H})}$.  $\partial{\bm{n}}/\partial{J_{AC}}$ is determined by the dynamics of the N\'eel vector and hence by the dynamics of sublattice magnetic moments, $\bm{m}_1$ and $\bm{m}_2$.

In the macrospin approximation, the free energy density (scaled into field dimensions) can be expressed in terms of the unit magnetization vectors $\bm{m}_{1,2}$ as
\begin{align}
\frac{\epsilon}{\hbar\gamma_{0}}=& H_\mathrm{ex}\bm{m}_{1} \cdot \bm{m}_{2} + H_{K}[(\bm{e}_{h}\cdot \bm{m}_{1})^2+(\bm{e}_{h}\cdot \bm{m}_{2})^2]\nonumber \\ &- H_{D}\bm{e}_D\cdot(\bm{m}_{1}\times\bm{m}_{2}) - \bm{H} \cdot (\bm{m}_{1}+\bm{m}_{2}),
\end{align}
where $\gamma_{0}=\mu_0\gamma$ is the product of the vacuum permeability $\mu_0$ and the gyromagnetic ratio $\gamma$. $\bm{e}_{h}$ and $\bm{e}_{D}$ are unit vectors in the directions of the hard axis anisotropy and DMI respectively. $H_\mathrm{ex}$, $H_{K}$ and $H_{D}$ are the effective fields associated with the exchange interaction, the hard axis anisotropy and the DMI, respectively.
Here we have ignored the in-plane easy axis anisotropy because it is much weaker than $H_{K}$ and $H_{D}$~\cite{Lebrun2019}. This simplification, however, becomes invalid when the in-plane projection of $\bm{H}$ is insufficient to maintain a single domain state~\cite{Cheng2019} (e.g., when $\bm{H}$ is close to being perpendicular to the easy plane). The dynamics of the sublattice magnetic moments are described by the coupled Landau-Lifshitz-Gilbert-Slonczewski equations

\begin{align}
\frac{d\bm{m}_{1,2}}{dt}=& \gamma_{0}\bm{H}_{1,2}^{\textrm{eff}}\times\bm{m}_{1,2} +\alpha_{0}\bm{m}_{1,2}\times\frac{d\bm{m}_{1,2}}{dt} \nonumber\\ &+ \gamma_{0}[H_{\textrm{Oe}}(J_\textrm{AC})+H_{\textrm{FL}}(J_\textrm{AC})] \bm{\sigma}_\textrm{AC} \times\bm{m}_{1,2} \nonumber \\ &+
\gamma_{0}H_{\textrm{DL}}(J_\textrm{AC})\bm{m}_{1,2}\times[\bm{m}_{1,2} \times \bm{\sigma}_\textrm{AC}],
\end{align}
where $\bm{H}_{1,2}^{\textrm{eff}}=-\delta\epsilon/(\hbar\gamma_{0}\delta\bm{m}_{1,2})$ is the effective field acting on $\bm{m}_{1,2}$, $\alpha_{0}$ is the Gilbert damping constant, $H_{\textrm{Oe}}(J_\textrm{AC})=\frac{1}{2}\mu_{0}A_{x}J_\textrm{AC}t_\mathrm{Pt}$ is the Oersted field and $t_\mathrm{Pt}$ the thickness of the Pt film. $\bm{\sigma}_\textrm{AC}$ is the direction of spin accumulation (here along $\hat{y}$), and $H_{\textrm{FL}}(J_\textrm{AC})$ and $H_{\textrm{DL}}(J_\textrm{AC})$ are the strengths of the field-like and damping-like torques, respectively.

All the three fields generated by the current are linearly proportional to $J_\textrm{AC}$. In particular, $H_{\textrm{Oe}}$ and $H_{\textrm{FL}}$ satisfy the same symmetry, but they are distinguishable by the dependence on the Pt film thickness. Because the intrinsic frequency of the spin dynamics in $\alpha$-Fe$_2$O$_3$ is several orders of magnitude larger than $\omega$~\cite{Seavey1972}, we can treat $\bm{m}_{1,2}$ as quasi-static vectors that adiabatically adjust to the AC current, remaining in the instantaneous ground state in the presence of current-induced torques. Under the adiabatic approximation, Eq. (3) can be solved analytically after a tedious derivation~\cite{Zhang2021}, which, if $\vert\bm{m}\vert=\vert\bm{m}_{1}+\bm{m}_{2}\vert \ll 1$ , can be expressed by the N\'eel vector $\bm{n}$ as function of $J_\textrm{AC}$ and $\bm{H}$. Finally, by inserting $\bm{n}(J_\textrm{AC},\bm{H})$ into Eq. (1), we obtain the general solution of SMR and the first two harmonic responses $V_{xy}^{1\omega}$ and $V_{xy}^{2\omega}$. This general solution, however, cannot be recast into a concise form unless we assume that  $\bm{e}_{h} \parallel \hat{z}$ and $\bm{e}_{D} \parallel \hat{z}$ i.e. the easy plane coincides with the film plane without tilting). The special solution for vanishing tilt is:

{\small
\begin{align}
R_{xy}^{1\omega}=&-\frac{1}{2}R_{0} \sin 2\phi_{H}, \\
R_{xy}^{2\omega}=&R_{0} \Bigg[-\frac{\cos2\phi_{H}\cos\phi_{H}}{H\sin\theta_{H}}(H_{\textrm{Oe}}+H_{\textrm{FL}})\  +\\
&\frac{2H_{\textrm{ex}}H\cos\theta_{H}\cos2\phi_{H}\sin\phi_{H}H_{\textrm{DL}}}{2({H_{D}}^{2}+2H_{\textrm{ex}}H_{K})H\sin\theta_{H}+H_{D}(H^{2}+4H_{\textrm{ex}}H_{K}+{H_{D}}^{2})}\Bigg]\nonumber,
\end{align}
}%
where $\theta_{H}\in[0,\pi]$ and $\phi_{H}\in[0,2\pi]$ are the polar and azimuthal angles of $\bm{H}$, and $R_{0}$ is a current-independent constant. In Fig. 2, we identify these angles as $\theta_{H}=\pi/2$  and $\phi_{H}=\alpha$ for the \textit{XY} scan, $\theta_{H}=\beta$ (or $2\pi - \beta$) and $\phi_{H}=0$ (or $-\pi$)  for the \textit{XZ} scan, and $\theta_{H}=\gamma$ (or $2\pi - \beta$) and $\phi_{H}=\pi/2$ (or $-\pi/2$) for the \textit{YZ} scan, respectively. To better fit the experimental data shown in Fig. 2, however, we further allow a small tilting of the hard axis but still assume that $\bm{e}_{D}$ $\parallel$ $\bm{e}_{h}$, which yields a complicated expression not shown here. We observe a good agreement with the experiment at a tilt angle of 3\textdegree\ with respect to $\hat{z}$, which is plotted by the red curves in Fig. 2. Since the model assumes $|\bm{m}|\ll 1$, our solution breaks down at extremely large fields, which is partially reflected in Fig. 3 and discussed later.

The field-like torque and the damping-like torque play very different roles in driving the dynamics of magnetic moments. As illustrated in Fig. \ref{Fig1:Measurement Geometry}(b), the damping-like torque cants both magnetic moments towards the same in-plane direction, which will subsequently leverage the exchange torque between $\bm{m}_{1}$ and $\bm{m}_{2}$ so that the N\'eel vector $\bm{n}$ develops an out-of-plane component. By contrast, the field-like torque acts as an effective field that directly drives the net magnetic moment so that the AC current induces an in-plane rotation of $\bm{m}$. Since $\bm{m} \perp \bm{n}$ by definition, a direct consequence is that $\bm{n}$ undergoes an in-plane oscillation. Correspondingly, in the absence of tilting, damping-like torques vanish in the \textit{XY} scan ($\theta_H=90$\textdegree), and field-like torques vanish in \textit{YZ} scan ($\phi_H=90$\textdegree). However, any amount of tilting of the easy plane will prevent this vanishing and both field-like and damping-like torques will have contributions in \textit{XY} and \textit{YZ} scans.

Because of the small magnetic moment present in $\alpha$-Fe$_2$O$_3$ above its Morin transition, the N\'eel vector couples to the external magnetic field perpendicularly. The easy plane and the external field direction together are adequate to uniquely set the equilibrium direction of the N\'eel vector and a $\gtrsim 1$~T in-plane component of the applied field can fully align the N\'eel vector to overcome any in-plane magnetic anisotropy.

\begin{figure}[ht]
  \centering
   \includegraphics[width=1\columnwidth]{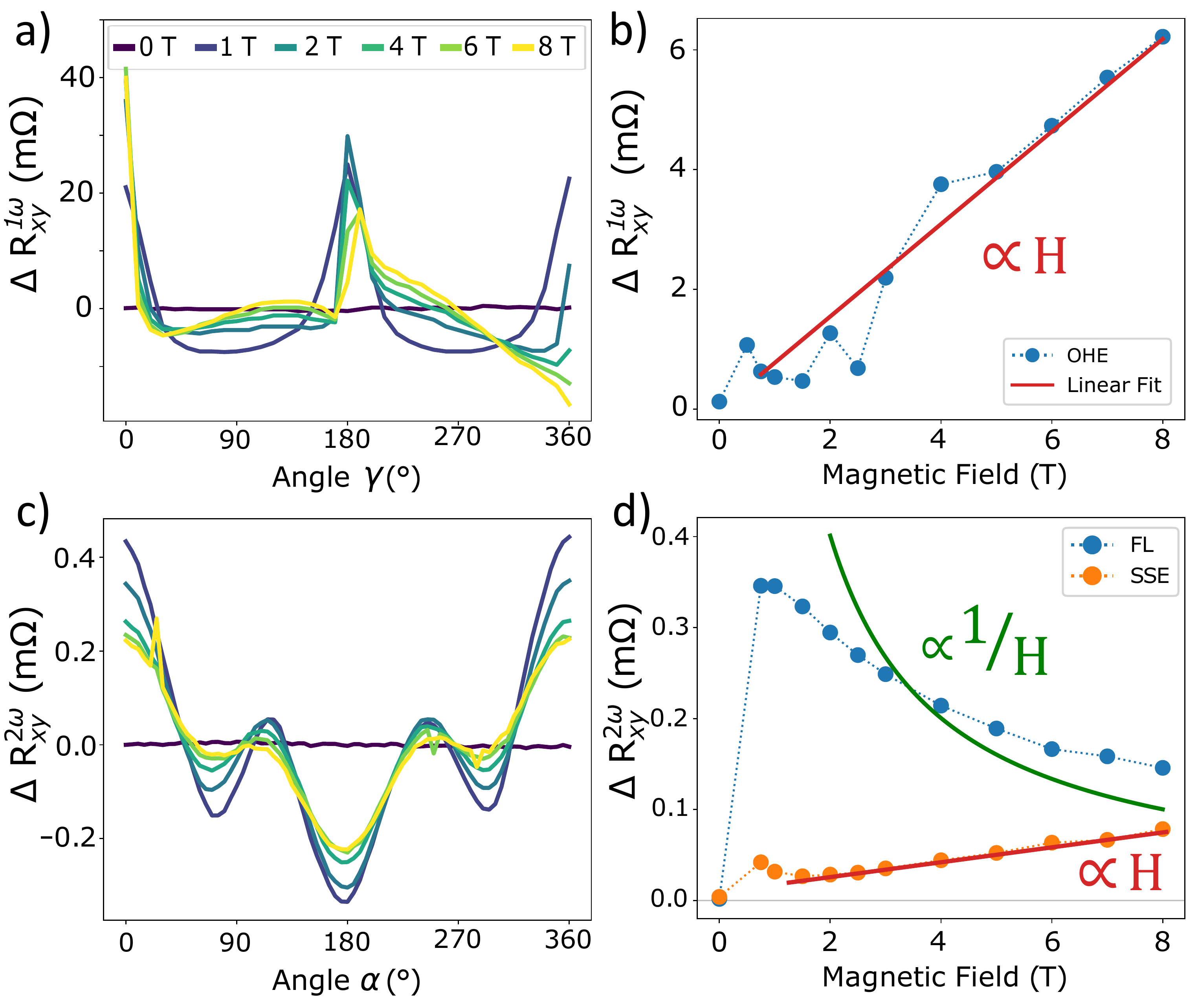} 
    \caption{Separating Hall effect and spin Seebeck effect contributions. a) Field dependence of $1^\mathrm{st}$ harmonic response in \textit{YZ} scan, from 0 to 8 T as indicated by the legend. b) The Hall effect in $1^\mathrm{st}$ harmonic \textit{YZ} scan (a), as a function of applied field. The linear trend line supports the claim that the origin of the signal is the ordinary Hall effect. c) Field dependence of $2^\mathrm{nd}$ harmonic response in \textit{XY} scan, which shares the same legend as (a). d) Antiferromagnetic spin Seebeck effect (SSE) separated from field-like SOT in $2^\mathrm{nd}$ harmonic \textit{XY} scan (c), as a function of applied field. The field-like component scales as $1/H$, whereas the SSE component scales linearly with $H$.}
\label{Fig3:OHE and SSE}
\end{figure}

We fit the experimental results with 3 free parameters: the direction of the hard axis ($\bm{e}_{h}$) and the amplitudes of the spin-orbit torques ($H_{\textrm{FL}}$ and $H_{\textrm{DL}}$). For every scan, first we fit the $1^\mathrm{st}$ harmonic response, where we extract the current-independent constant $R_{0}$ (Eq. 4). This resistance is necessary to correctly extract other desired parameters from the second harmonic response (Eq. 5). Then together with the $R_{0}$, we use material parameters from the literature $H_{D}=$2 T, $H_{K}=$ 0.01 T and $H_{\textrm{ex}}=900$ T \citep{Williamson1964,Mizushima1966,Elliston1968} to fit the responses with our model. These fits allow us to extract the amplitudes of effective fields associated with the spin-orbit torques per current density which are $H_{\textrm{FL}}/J_{\textrm{AC}}\approx 7.5 \times $10$^{-2}\ $T$/(10^{12}$A/m$^{2}$) and $H_{\textrm{DL}}/J_{\textrm{AC}}\approx 4.2  \times $10$^{-4}\ $T$/(10^{12}$A/m$^{2}$).

A slight tilting ($\sim$ 3\textdegree) of the hard-axis $\bm{e}_h$ and $\bm{e}_D$ (again, $\bm{e}_D\parallel\bm{e}_h$ in our model) from the $\hat{z}$-direction is needed in order to simultaneously capture the form of each and every one of the field scans. This tilt is especially crucial in 1$^\mathrm{st}$ harmonic out-of-plane scans (\textit{XZ} and \textit{YZ}), where in order to get any non-zero response from our model, we need some degree of canting of the easy plane. Moreover, to explain the shape of the response, an additional cosine term is needed, which we interpret as the ordinary Hall effect response of Pt.

In our experiments, we rotate the applied magnetic field both in the film plane and out of the film plane while passing high current densities across the Pt leads. Therefore, there will be effects arising from the ordinary Hall effect induced by the perpendicular field component and from the spin Seebeck effect (SSE) due to a thermal gradient in the $\hat{z}$ direction caused by Joule heating. Figure~\ref{Fig3:OHE and SSE} summarizes how we extracted these contributions. On the left (Fig.~\ref{Fig3:OHE and SSE}(a) and (c)), we show the field dependence of 1$^\mathrm{st}$ harmonic YZ and 2$^\mathrm{nd}$ harmonic XY scans. And, on the right (Fig.~\ref{Fig3:OHE and SSE}(b) and (d)), we show the separated contributions of Hall effect and SSE from those scans, respectively. The field dependence of the signals can be used to isolate different contributions to the signal, as well as to show the model field dependence with the experimental results. 

\begin{figure}[t]
  \centering
   \includegraphics[width=1\columnwidth]{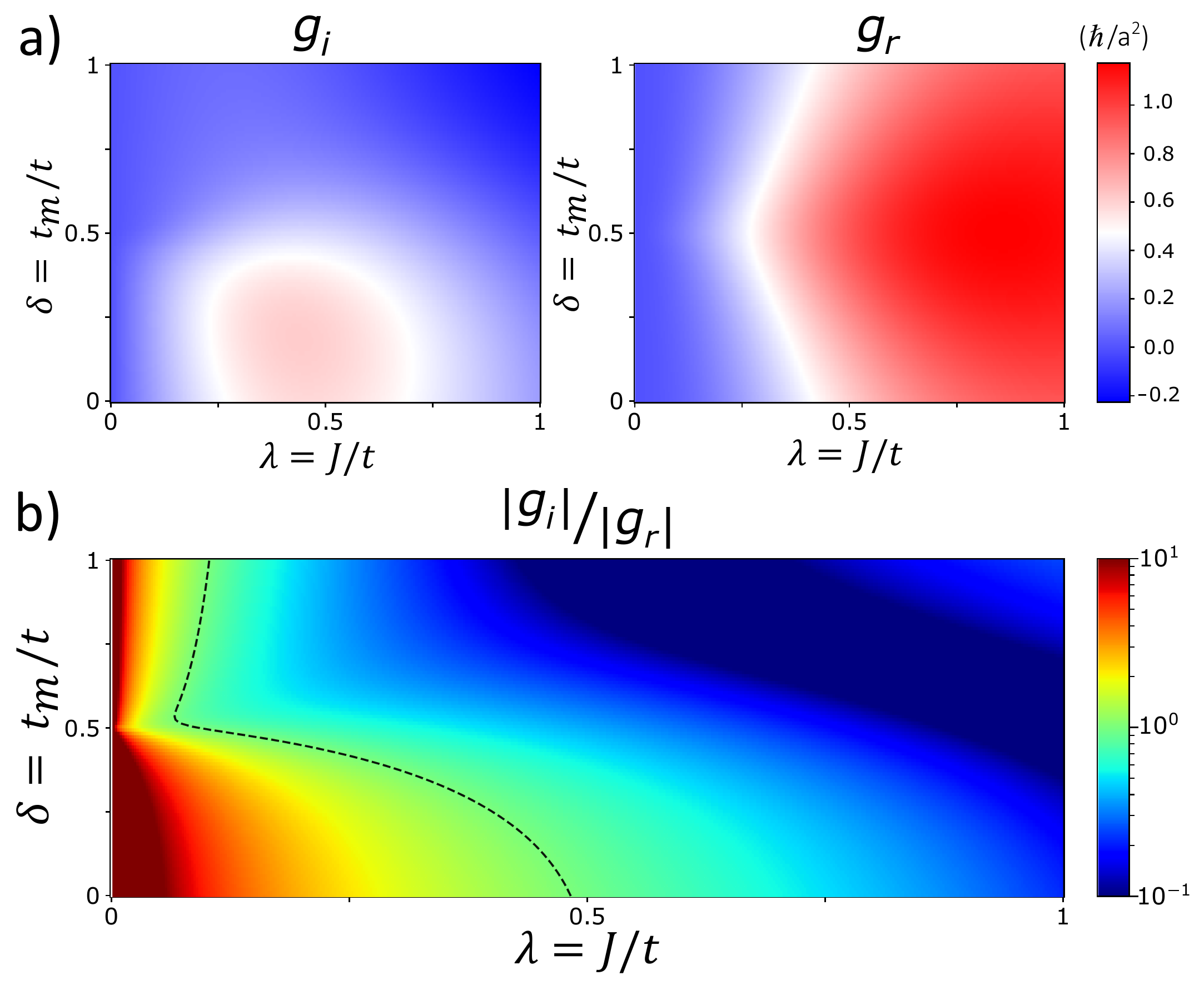} 
    \caption{a) Imaginary (left) and real (right) parts of the spin mixing conductance ($g = g_{r}+ig_{i}$) of an AFM/NM interface. The scale is indicated by the shared color bar on the right, with units of $\hbar$ per unit area $a^2$ on the interface, where $a$ is the lattice constant. b) Ratio of $|g_i|$ to $|g_r|$. As $\lambda$ approaches zero $|g_i|$ dominates over $|g_r|$. The dashed line shows $|g_i|=|g_r|$, which separates the field-like torque dominating region ($g_{i}>g_{r}$) on the left from the damping-like torque dominating region ($g_{r}>g_{i}$) on the right. The color bar scale is logarithmic in this case.}
\label{Fig4:Spin Mixing Conductance}
\end{figure}

Figure~\ref{Fig3:OHE and SSE}(b) shows the Hall effect contribution in 1$^\mathrm{st}$ harmonic and \textit{YZ} scans. Since the expected Hall contribution is in the form $R^{1\omega}_{H}=H R^{H}_{0}\cos{\theta}$, the linear trend with applied field supports our claim. Similarly, Fig.~\ref{Fig3:OHE and SSE}(d) shows decomposition of $R_{xy}^{2\omega}$ into two components: Field-like SOT (blue) $R_{\textrm{FL}}^{2\omega}$ and spin Seebeck effect (orange) $R_{\textrm{SSE}}^{2\omega}$. In this case, the expected signals are of the form $R^{2\omega}_\textrm{FL}=HR^\textrm{FL}_0\cos{2\alpha}\cos{\alpha}$ and $R^{2\omega}_\textrm{SSE}=H R^\textrm{SSE}_0\cos{\alpha}$ for field-like and SSE contributions, respectively. 
The form of the SSE contribution indicates that the effect is associated with the excess moment in the so called the transverse spin Seebeck geometry, in which the signal scales linearly with the applied field \citep{Bauer2012}. Furthermore, the remaining field-like component decreases with increasing applied magnetic field and follows a $1/H$ dependence, as in Eq.~5. We attribute the deviation from a $1/H$ dependence at low and high fields to the fact that our model assumptions are not valid in those limits. 

Our experimental results show that the field-like torque is about two orders of magnitude larger than the damping-like torques. This implies that in the spin mixing conductance $g$ ($g = g_{r}+ig_{i}$)~\cite{Cheng2014}---which characterizes the interaction between itinerant electrons in the Pt and spins in the AFM---the imaginary component $g_i$ far exceeds the real part $g_{r}$. This follows from  the relation $H_\textrm{DL}= J_\textrm{AC} (a^3/ed_{M})g_r f_d$ and $H_\textrm{FL}= J_\textrm{AC} (a^3/ed_{M})g_i f_d$ (see details in Supplementary Materials). The finding $g_{i} \gg g_{r}$ is contrary to that of ferromagnetic/nonmagnetic heterostructures~\cite{tserkovnyak2002spin,xia2002spin}. To better understand this counterintuitive result, we calculate $g_i$ and $g_r$ through the interfacial spin-dependent scattering~\cite{cheng2014aspects}, where the scattering processes depend on three parameters: the electron hopping energy $t$ in the NM (determined by the Fermi energy), the proximity-induced hopping $t_m$ on the AFM side, and the interfacial exchange coupling $J$ between conduction electrons in the NM and the magnetic moments in the AFM. By spin-flip scattering, electrons can deliver angular momenta to the AFM which manifests as spin torques exerting on the magnetic moments. In AFM, the Umklapp scattering is essential, which involves communications between the two magnetic sublattices, hence relying on $t_{m}$. In Fig.~\ref{Fig4:Spin Mixing Conductance} we plot the dependence of $g_{r}$ and $g_{i}$ on two dimensionless parameters $\lambda = J/t$ and $\delta = t_{m}/t$. Details on the calculations are discussed in the section 4 of the Supplementary Materials.

We see from Fig.~\ref{Fig4:Spin Mixing Conductance} that regions with small $\lambda$ and $\delta$ have $g_i/g_r>1$. This ratio can become extremely large when $\lambda$ becomes even smaller (see Supplementary Materials), which is consistent with our experimental results. Because the ratio $g_i/g_r$ can vary over a wide range of values depending on $\lambda$ and $\delta$, it is expected that
the relative strength of the field-like and damping-like torques can vary significantly in different materials. It should be noted that the conservation of angular momentum requires $g_{r} > 0$, and $g_{i} < 0$ does not violate this requirement (see the definitions of $g_{r}$ and $g_{i}$ in the Supplementary Materials). We also mention that in recent spin pumping experiments~\cite{vaidya2020subterahertz,li2020spin,wang2021spin,boventer2021room}, only $g_r$ could be determined because the DC component of the pumped spin current only depends on $g_r$, whereas $g_i$ only contributes to the AC spin pumping. Our experiment, on the other hand, can directly probe the effect of $g_i$.

In conclusion, we determined the type and amplitude of SOTs in antiferromagnetic $\alpha$-Fe$_2$O$_3$/Pt heterostructures by performing harmonic Hall measurements. The model we developed suggests that both in-plane and out-of-plane scans are important to accurately characterize the nature of SOTs. We also found that field-like torques are two orders of magnitude larger than damping-like torques, contrary to what has been reported in similar structures with ferromagnets. This implies that the spin-mixing conductance of the $\alpha$-Fe$_2$O$_3$/Pt interface has the unusual property of having a large imaginary component. Our work demonstrates a straightforward way to quantify SOTs and opens up a promising path for future studies on similar AFM/HM heterostructures as well as a means that can be used in optimizing SOT on AFM for applications.

\section*{Acknowledgements}

This research was supported by the Air Force Office of Scientific Research under Grant FA9550- 19-1-0307. The nanostructures were realized at the Advanced Science Research Center NanoFabrication Facility of the Graduate Center at the City University of New York.

\end{document}